\documentclass[fleqn,usenatbib]{mnras}

\usepackage{newtxtext,newtxmath}

\usepackage[T1]{fontenc}

\DeclareRobustCommand{\VAN}[3]{#2}
\let\VANthebibliography\thebibliography
\def\thebibliography{\DeclareRobustCommand{\VAN}[3]{##3}\VANthebibliography}

\usepackage{graphicx}	
\usepackage{amsmath}	
\usepackage[version=3]{mhchem} 
\usepackage{siunitx} 

\usepackage{mathabx} 
\usepackage[dvipsnames]{xcolor}
\usepackage{ulem}

\DeclareSIUnit\year{yr}
\DeclareSIUnit\parsec{pc}
\DeclareSIUnit\sterradian{sr}
\DeclareSIUnit\erg{erg}

\newcommand{\stand}{S{\footnotesize{TAND}2019}}
\newcommand{\standolder}{S{\footnotesize{TAND}2015}}
\newcommand{\argo}{A{\footnotesize{RGO}}}

\title[MOVES IV]{MOVES IV. Modelling the influence of stellar XUV-flux, cosmic rays, and stellar energetic particles on the atmospheric composition of the hot Jupiter HD~189733b}

\author[P. Barth et al.]{P. Barth,$^{1,2,3}$\thanks{E-mail: pb94@st-andrews.ac.uk}
        Ch. Helling,$^{1,2,4}$
        E. E. Stüeken,$^{1,3}$
        V. Bourrier,$^{5}$
        N. Mayne,$^{6}$
        P. B. Rimmer,$^{7,8,9}$
        \newauthor{M. Jardine,$^{2}$
        A. A. Vidotto,$^{10}$
        P. J. Wheatley,$^{11,12}$
        R. Fares$^{13}$}
\\
$^{1}$Centre for Exoplanet Science, University of St Andrews, North Haugh, St Andrews, KY169SS, UK \\
$^{2}$SUPA, School of Physics \& Astronomy, University of St Andrews, North Haugh, St Andrews, KY169SS, UK \\
$^{3}$School of Earth \& Environmental Sciences, University of St Andrews, St Andrews, Fife, KY16 9AL, UK \\
$^{4}$SRON Netherlands Institute for Space Research, Sorbonnelaan 2, 3584 CA Utrecht, NL \\
$^{5}$Observatoire de l’Université de Genève, Chemin des Maillettes 51, Versoix, CH-1290, Switzerland \\
$^{6}$Physics and Astronomy, College of Engineering, Mathematics and Physical Sciences, University of Exeter, EX4 4QL, UK \\
$^{7}$Department of Earth Sciences, University of Cambridge, Downing St, Cambridge CB2 3EQ, UK \\
$^{8}$Cavendish Astrophysics, JJ Thomson Ave, Cambridge CB3 0HE, UK \\
$^{9}$MRC Laboratory of Molecular Biology, Francis Crick Ave, Cambridge CB2 0QH, UK \\
$^{10}$School of Physics, Trinity College Dublin, College Green, D-2, Ireland \\
$^{11}$Centre for Exoplanets and Habitability, University of Warwick, Gibbet Hill Road, Coventry CV4 7AL, UK \\
$^{12}$Department of Physics, University of Warwick, Gibbet Hill Road, Coventry CV4 7AL, UK \\
$^{13}$Physics Department, United Arab Emirates University, P.O. Box 15551, Al-Ain, United Arab Emirates \\
}

\date{Accepted 2020 December 22. Received 2020 December 22; in original form 2020 September 18}

\pubyear{2020}

\begin{document}
\label{firstpage}
\pagerange{\pageref{firstpage}--\pageref{lastpage}}
\maketitle

\begin{abstract}
{Hot Jupiters provide valuable natural laboratories for studying potential contributions of high-energy radiation to prebiotic synthesis in the atmospheres of exoplanets.}
{In this fourth paper of the MOVES (Multiwavelength Observations of an eVaporating Exoplanet and its Star) programme, we study the effect of different types of high-energy radiation on the production of organic and prebiotic molecules in the atmosphere of the hot Jupiter HD 189733b.}
{Our model combines X-ray and UV observations from the MOVES programme and 3D climate simulations from the 3D Met Office Unified Model to simulate the atmospheric composition and kinetic chemistry with the STAND2019 network.
Also, the effects of galactic cosmic rays and stellar energetic particles are included.}
{We find that the differences in the radiation field between the irradiated dayside and the shadowed nightside lead to stronger changes in the chemical abundances than the variability of the host star's XUV emission.
We identify ammonium (\ce{NH4^+}) and oxonium (\ce{H3O^+}) as fingerprint ions for the ionization of the atmosphere by both galactic cosmic rays and stellar particles.
All considered types of high-energy radiation have an enhancing effect on the abundance of key organic molecules such as hydrogen cyanide (\ce{HCN}), formaldehyde (\ce{CH2O}), and ethylene (\ce{C2H4}).
The latter two are intermediates in the production pathway of the amino acid glycine (\ce{C2H5NO2}) and abundant enough to be potentially detectable by \textit{JWST}.}
{}

\end{abstract}




\section{Introduction}

HD~189733b \citep{Bouchy2005} orbits an active K dwarf (HD~189733~A, hereafter HD~189733) at a distance of $\SI{0.031}{\astronomicalunit}$ and is the closest transiting hot Jupiter to Earth and particularly well studied \citep[e.g.][and references therein]{Helling2016,Lines2018,Drummond_2020,Odert2020}.
Various molecules have been detected in the planet's atmosphere so far, including \ce{H2O} \citep{Grillmair2008,Birkby2013,Crouzet2014,Todorov2014,McCullough2014}, \ce{CH4} \citep{Swain2014}, and \ce{CO} \citep{Desert2009,DeKok2013,Rodler2013}.
However, thick clouds and photochemical hazes obscure the deeper layers of the atmosphere where other molecules are expected to be present \citep{Gibson2012,Deming2013,Pont2013,Helling2016,Iyer2016,Zahnle2016,Kirk2017}.
Characterization of these deeper atmospheric layers therefore requires computational models that take into account various types of radiation and gas transport mechanisms.
Such simulations are particularly important for unravelling the role of atmospheric processes in prebiotic reactions that may potentially lead to the establishment of a living biosphere.
Although life is not expected to originate on hot Jupiters, the wealth of observational data collected from these objects over recent years make them ideal natural laboratories for improving our understanding of fundamental gas phase reactions on extrasolar worlds.

In previous work, \citet{Moses2011} carried out photo- and thermochemical kinetics simulations of the atmosphere of HD~189733b with a neutral H/C/N/O network, showing an enhancement of \ce{NH3} and \ce{CH4} due to transport-induced quenching in the middle layers of the atmosphere.
In the upper atmosphere, photochemistry enhances the abundances of atomic species, unsaturated hydrocarbons, nitriles, and radicals.
More recent studies by \citet{Venot2012a,Venot2020} with a chemical kinetics network that is validated by experimental data showed differences in the abundance of \ce{NH3}, \ce{HCN}, and \ce{CH4} due to different quenching pressures.
Some work has also been done to investigate the effects of high-energy radiation.
For example, \citet{Venot2016} studied the influence of stellar flares on the chemical composition of exoplanets at the example of a sub-Neptune/super-Earth-like planet orbiting an active M star.
They found that individual and recurring flares can irreversibly alter the abundances of key molecules such as H or \ce{OH}.
However, in their simulations they only considered flares in the stellar radiation and not the effect of charged particles which could be associated with a stellar flare.
In contrast, \citet{Segura2010} included the effect of stellar particles into their simulations to study the effect of flares on the oxygen-rich atmosphere of an Earth-like planet orbiting an active M star in its habitable zone.
They found that such flares can severely deplete the ozone layer over several decades, which illustrates the importance of including energetic particles in realistic atmospheric chemistry models.
\citet{Chadney2017} studied the effect of stellar flares on the upper atmosphere ($p < 10^{-6} \si{\bar}$) of HD~189733b and HD~209458b where they found an enhancement in the electron densities of a factor of 2.2 -- 3.5.
However, while this study provided an estimate on the mass loss rate due to a stellar proton event, and also took into account photoionisation and electron-impact ionisation of H and He, the effects of stellar energetic protons on atmospheric chemistry were not explored.

This paper is part of the MOVES collaboration (\textit{Multiwavelength Observations of an eVaporating Exoplanet and its Star}, PI V.~Bourrier) which aims to characterize the complex and variable environment of the HD~189733 system.
In MOVES I, \citet{Fares2017} studied the evolving magnetic field of the host star HD~189733, an active K~dwarf.
They used spectropolarimetric data to reconstruct the magnetic field of the star and study its evolution over nine years.
\citet{Kavanagh2019} (MOVES II) used these results to perform 3D MHD simulations of the stellar wind of HD~189733 and to predict the radio emission from both the planet and the star.
The models predict that the planet experiences a non-uniform wind with the wind velocity and particle number density varying by up to 37~\% and 32~\%, respectively.
\citet{Bourrier2020} (MOVES III) confirmed the variable HI escape from HD~189733b \citep{LecavelierdesEtangs2012,Bourrier2013}, and used \textit{HST}, \textit{XMM-Newton}, and \textit{Swift} data to derive semi-synthetic XUV spectra of the host star that we use for our simulations (Section~\ref{SubSec_XUV}).

In our study, we use these semi-synthetic XUV spectra to investigate the effect of stellar X-ray and UV (XUV) radiation, as well as the effects of stellar energetic particles (SEP) and cosmic rays (CR) on the atmospheric chemistry of HD~189733b.
First, in Section~\ref{Sec_Network} we introduce the chemical kinetics network \stand{} \citep{Rimmer2016, Rimmer2019b, Rimmer2019} which we used for our simulations.
In Section~\ref{Sec_Input} we discuss the atmospheric pressure-temperature profiles and the parameterization of the different sources of high-energy radiation before we present the resulting abundances of neutral and charged molecules at the substellar and antistellar point (Section~\ref{Sec_Results}).
We put particular emphasis on the effect of high-energy radiation sources on the ionization of the atmospheric gas as precursor for the formation of a magnetosphere.
Furthermore, we explore the possible production of prebiotic organic molecules in giant gas planets as the atmosphere of HD~189733b provides a potentially observable laboratory for the onset of prebiotic synthesis.

\section{Approach}
\label{Sec_Network}

\begin{table*}
    \caption{New reactions added to \protect\stand{}. The full network is available in the supplementary materials. Note that reverse reactions are not included in the network file.}
	\begin{tabular}{ccccccc}
		\noalign{\smallskip}
		\hline
		\noalign{\smallskip}
		Number & Reaction & Class$^c$ & $\alpha$ & $\beta$ & $\gamma$ & Source \\
		\noalign{\smallskip}
		\hline \hline
		\noalign{\smallskip}
		627$^l$ & \ce{Cl^+ + e^- -> Cl} & T & $3.43\times10^{-14}$ & -3.77 & 0 & Reaction no. 623$^a$ \\
		628$^h$ & \ce{Cl^+ + e^- -> Cl} & T & $10^{-7}$ & 0 & 0 & Reaction no. 624$^a$ \\
		1406 & \ce{C2N + H -> HCN + C} & A & $2\times10^{-10}$ & 0 & 0 & \citet{Loison2014} \\
		1407 & \ce{CNC + H -> HCN + C} & A & $2\times10^{-10}$ & 0 & 0 & Reaction no. 1406 \\
		3072 & \ce{C3^+ + e^- -> C2 + C} & U & $3\times10^{-7}$ & -0.5 & 0 & OSU$^b$, \citet{Loison2017} \\
		3085 & \ce{Cl^+ + e^- -> Cl} & RA & $1.13\times10^{-10}$ & -0.7 & 0 & OSU$^b$ \\
		\noalign{\smallskip}
		\hline
	\end{tabular}\\
	$^c$ reaction classes: T -- Three-body recombination reactions, A -- Neutral bimolecular reactions, U -- Dissociative recombination reactions, RA -- Radiative association reaction, $^l$ low pressure regime, $^h$ high pressure regime, $^a$ rates adopted from corresponding reaction with \ce{Na^+}, $^b$ Ohio State University chemical network \citep{Harada2010}.
	\label{Tab_New_Reac}
\end{table*}

We use the chemical kinetics network \stand{} \citep{Rimmer2016, Rimmer2019b, Rimmer2019} to simulate the influence of XUV radiation and cosmic rays on the atmospheric chemistry.
\stand{} is an H/C/N/O network with reactions involving \ce{He}, \ce{Na}, \ce{Mg}, \ce{Si}, \ce{Cl}, \ce{Ar}, \ce{K}, \ce{Ti}, and \ce{Fe}.
It contains all reactions for species of up to six H, two C, two N, and three O atoms and is valid for temperatures between 100 and $\SI{30000}{\kelvin}$.
The network has been benchmarked against modern Earth and Jupiter \citep{Rimmer2016}, as well as hot-Jupiter models \citep{Tsai2017,Hobbs2019}.
The full network is available in the supplementary materials.

\stand{} uses a 1D photochemistry and diffusion code \argo{} which requires the following inputs:
\begin{enumerate}
    \item $p_{\rm gas}-T_{\rm gas}$ profile of the atmosphere
    \item Vertical eddy diffusion coefficient profile $K_{zz}$
    \item Atmospheric element abundances
    \item Boundary conditions at the top and bottom of the atmosphere
    \item Actinic flux at the top of the atmosphere
    \item Chemical network (\stand{})
    \item Initial chemical composition
\end{enumerate}
\argo{} follows a single gas parcel while it moves up- and downwards through the atmosphere due to eddy diffusion. The kinetic gas-phase chemistry has no feedback onto the initial ($T_{\rm gas}$, $p_{\rm gas}$)-profiles.

The model consists of two parts: a chemical transport model that considers ion-neutral and neutral-neutral reactions and a model calculating  the chemical rate constants for photochemistry and cosmic rays.
The gas-phase number densities, $n_{\rm i}$, are determined by 1D continuity equations \citep{Rimmer2016,Rimmer2019b},
\begin{equation}
    \frac{\partial n_i}{\partial t} = P_i - L_i - \frac{\partial \Phi_i}{\partial z},
    \label{eq:rates}
\end{equation}
where $n_i$ is the number density of species $i$, $P_i$ is the rate of production and $L_i$ the rate of loss of species $i$ (both in $\si{\per\centi\metre\cubed\per\second}$). $\frac{\partial \Phi_i}{\partial z}$ is the vertical change in the flux $\Phi_i [\si{\per\centi\metre\squared\per\second}]$ due to Eddy and molecular diffusion.
Ambipolar diffusion is not included in the model.
In the presence of a magnetic field, this type of diffusion might slow down vertical mixing but as long as the degree of ionisation remains low the effect on the chemistry is not expected to be large.
For strongly ionized atmospheres, ambipolar diffusion can be parameterized to represent the electron-induced electric field that is acting on the ions.

The source and sink term $P_i$ and $L_i$ in Eq.~(\ref{eq:rates}) describe two-body neutral-neutral and ion-neutral reactions, three-body neutral reactions, dissociation reactions, radiative association reactions, as well as thermal ionization and recombination reactions.
In total, the version of \stand{} used in this study includes 3085 forward reactions with 197 neutral and 224 ion species and is included in the supplementary material (note that the reverse reactions are not included in the available network file).
The network is a modified version of \stand{} \citep{Rimmer2019b,Rimmer2019} which itself is a new version of the original \standolder{} network \citep{Rimmer2016}.
In the following, we will describe the changes made to the network for this work which are summarised in Table~\ref{Tab_New_Reac}.

\subsection{Additions to the network}

One of the cosmic ray ionization reactions in \stand{} leads to the production of \ce{Cl^+}, but until now the network has not included any destruction reactions for \ce{Cl^+}, leading to an overestimation of the abundance of this ion. We therefore have added additional production and destruction reactions for \ce{Cl^+}:
\begin{eqnarray}
    \label{Reac_1351}
    \ce{Cl + M &->& Cl^+ + e^- + M}, \\
    \label{Reac_1193}
    \ce{Cl^+ + e^- + M &->& Cl + M},
\end{eqnarray}
where M represents a not defined third body.
As there are no published rate coefficients for these reactions, we used the rates for the corresponding \ce{Na} reactions.
Even though \ce{Na} and \ce{Cl} have different electron affinities, that should not strongly affect the reaction rates.
We will make sure to update these coefficients as soon as there is new data available.

The rates of these thermal ionization (Eq.~\ref{Reac_1351}) and three-body recombination reactions (Eq.~\ref{Reac_1193}) follow the Lindemann form \citep{Lindemann1922}, which provides pressure dependent reaction rates with limits for the low ($k_0 [\si{\centi\metre^6\per\second}]$) and the high-pressure regimes ($k_\infty [\si{\centi\metre^3\per\second}]$):
\begin{eqnarray}
    k_0      &=& \alpha_0      \left( \frac{T}{\SI{300}{\kelvin}} \right)^{\beta_0}      e^{-\gamma_0/T}, \\
    k_\infty &=& \alpha_\infty \left( \frac{T}{\SI{300}{\kelvin}} \right)^{\beta_\infty} e^{-\gamma_\infty/T}.
\end{eqnarray}
The coefficients $\alpha_i$, $\beta_i$, and $\gamma_i$ for above rates can be found in Table~\ref{Tab_New_Reac} (reactions 1193 and 1351 for the low-pressure and reactions 1194 and 1352 for the high-pressure regime).
The pressure-dependent rate can now be calculated with the reduced pressure $p_\mathrm{r} = k_0 n_\mathrm{M}/k_\infty$, where $n_\mathrm{M} [\si{\per\centi\metre\cubed}]$ is the number density of the neutral third body \citep{Rimmer2016}:
\begin{equation}
    k = \frac{k_\infty p_\mathrm{r}}{1 + p_\mathrm{r}}.
\end{equation}

In addition, a radiative association reaction for the destruction of \ce{Cl^+} (Reaction 5742) was added:
\begin{equation}
    \ce{Cl^+ + e^- -> Cl +} h \nu,
\end{equation}
where the excess energy of the recombination is emitted as a photon with energy $h\nu$.
The reaction rate follows the Kooji equation \citep{Kooij1893}:
\begin{equation}
    \label{Eq_Kooij}
    k_\mathrm{RA} = \alpha \left( \frac{T}{\SI{300}{\kelvin}} \right)^\beta e^{-\gamma/T}.
\end{equation}
The coefficients (Table~\ref{Tab_New_Reac}) are taken from the KIDA database \citep{Wakelam2012} and originally from the Ohio State University (OSU) chemical network \citep{Harada2010}.
This rate is then combined with the rate for the three-body recombination reaction presented above (Eq.~\ref{Reac_1193}) to
\begin{equation}
    k = \frac{(k_0 n_\mathrm{M} F + k_\mathrm{RA}) k_\infty}{k_0 n_\mathrm{M} + k_\infty},
\end{equation}
where $F$ is a dimensionless function to more accurately approximate the transition from high to low pressure which gives the Troe form \citep{Troe1983,Rimmer2016}.

When running \argo{} for HD~189733b, HCCN was overproduced due to the absence of the destruction reactions for CNC. This was resolved via the addition of two neutral bimolecular reactions to the network:
\begin{eqnarray}
    \label{Reac_2208}
    \ce{C2N + H &->& HCN + C} \\
    \label{Reac_2209}
    \ce{CNC + H &->& HCN + C}
\end{eqnarray}
The coefficients for the first reaction (Eq.~\ref{Reac_2208}) were taken again from the KIDA database and were presented previously by \citet{Loison2014}.
For the destruction reaction of \ce{CNC} we assume the same rate constant.
Since $\beta$ and $\gamma$ are zero (Table~\ref{Tab_New_Reac}), the reaction rate is given by
\begin{equation}
    k = \alpha \left( \frac{T}{\SI{300}{\kelvin}} \right).
\end{equation}

Similar to the above described case of \ce{Cl^+}, destruction reactions for \ce{C3^+} were missing in \stand{}.
Therefore, a dissociative recombination reaction has been added to the network (Reaction 5729):
\begin{equation}
    \ce{C3^+ + e^- -> C2 + C}
\end{equation}
The reaction rate follows the Kooji form (Eq.~\ref{Eq_Kooij}).
The coefficients presented in Table~\ref{Tab_New_Reac} are again from the KIDA database and originally from the OSU network and \citet{Loison2017}.
Furthermore, the NASA coefficients for \ce{K^+} were updated to those from the Burcat catalogue \citep{Burcat2005}\footnote{http://garfield.chem.elte.hu/Burcat/burcat.html}

\section{Input}
\label{Sec_Input}

\subsection{Atmospheric profiles}
\label{SubSec_Profiles}

\begin{table}
    \caption{Model parameters for HD~189733b, adapted from \citet{Southworth2010} \citep[for more details see][Table 1]{Lines2018}}
	\begin{tabular}{lc}
		\noalign{\smallskip}
		\hline
		\noalign{\smallskip}
		Quantity & Initial value\\
		\noalign{\smallskip}
		\hline \hline
		\noalign{\smallskip}
		Specific gas constant$^a$, $R \; (\si{\joule\per\kilogram\per\kelvin})$ & 3556.8 \\
		Specific heat capacity, $c_\mathrm{p} \; (\si{\joule\per\kilogram\per\kelvin})$ & $1.3 \times 10^4$ \\
		Radius, $R_\mathrm{p} \; (\si{\metre})$ & $8.05 \times 10^7$\\
		Surface gravity, $g_\mathrm{p} \; (\si{\metre\per\second})$ & 22.49 \\
		Semi-major axis, $a_\mathrm{p} \; (\si{au})$ & $3.14 \times 10^{-2}$ \\
		\noalign{\smallskip}
		\hline
	\end{tabular}\\
	$^a$ Unlike noted in \citet[Tab.~1]{Lines2018} the specific gas constant is not $R = \SI{4593}{\joule\per\kilogram\per\kelvin}$ but $\SI{3556.8}{\joule\per\kilogram\per\kelvin}$ which was also used in the simulations in that paper and used by \citep{Drummond_2018b}.
	\label{Tab_Input_UM}
\end{table}

\begin{figure*}
    \centering
    \includegraphics[width=\textwidth]{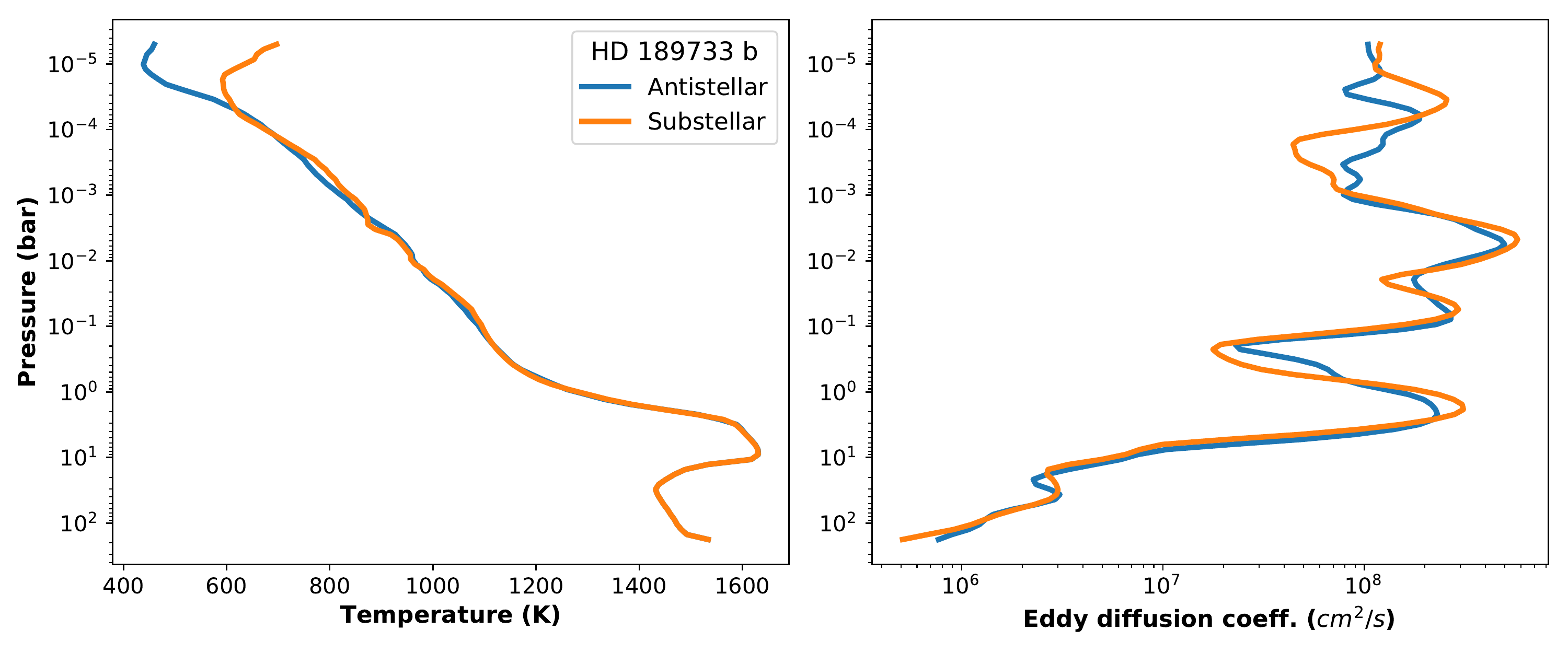}
    \caption{\textit{Left:} Atmospheric ($T_{\rm gas}, p_{\rm gas}$)-profiles for HD~189733b from simulations with the 3D Met Office UM \citep{Lines2018} at the antistellar ($\phi=0^\circ,\theta=0^\circ$) \& substellar point ($180^\circ,0^\circ$). \textit{Right:} Profiles of the eddy diffusion coefficient $K_\mathrm{zz}$ derived from the standard deviation of the vertical velocity.}
    \label{profiles_kzz_HD189733b}%
\end{figure*}

\begin{figure*}
    \centering
    \includegraphics[width=\textwidth]{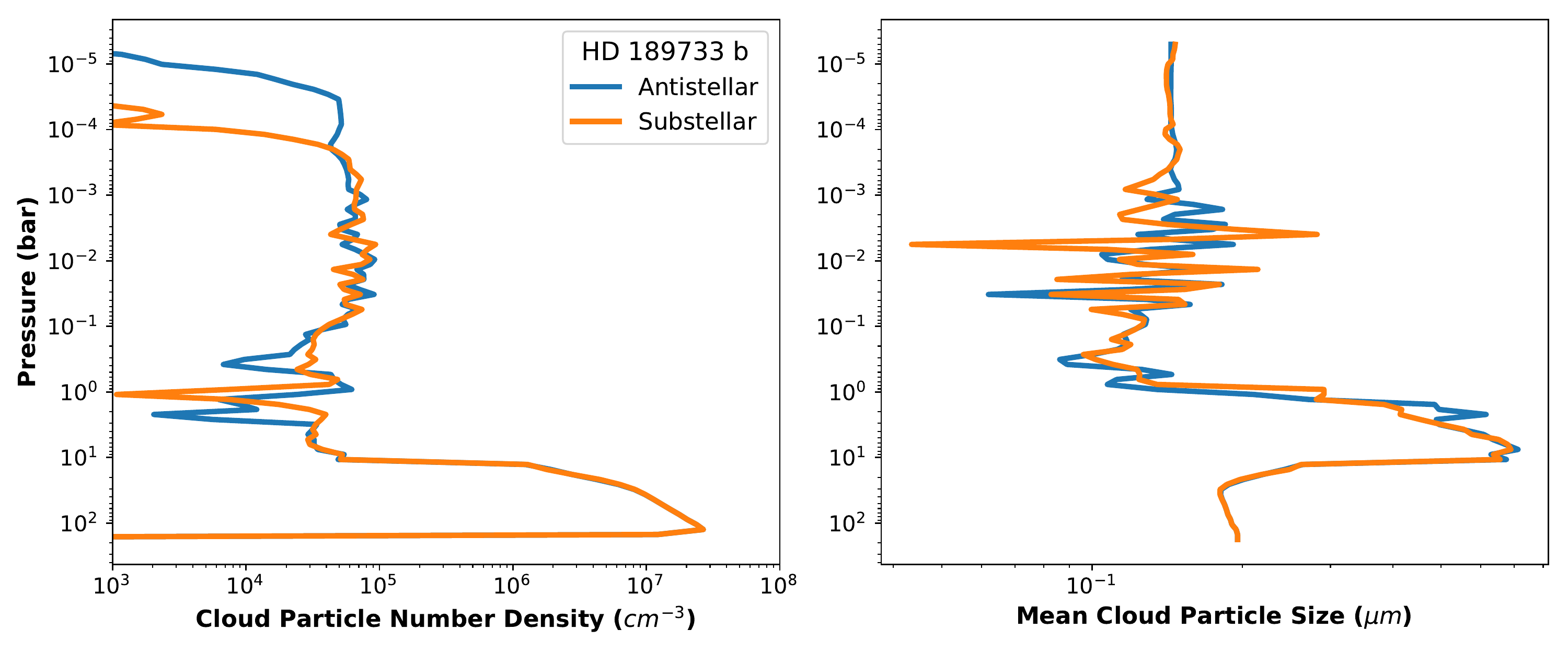}
    \caption{Cloud particle number density (\textit{left}) and mean cloud particle size (\textit{right}) for HD~189733b at the same profiles as in Fig. \ref{profiles_kzz_HD189733b}. Data extracted from simulations with the 3D Met Office UM \citep{Lines2018}.}
    \label{Cloud_numdens_size_HD189733b_new}%
\end{figure*}

We extract 1D profiles from the 3D Met Office Unified Model (UM) simulations of the atmosphere of HD~189733b by \citet{Lines2018} which consistently solve a kinetic cloud formation model \citep{Helling2001,Woitke2003,Woitke2004,Helling2006,Helling2008}.
Some of the input parameters for the UM are given in Table~\ref{Tab_Input_UM}.
Cloud properties like  particles sizes or their number density are consistent with the thermodynamic profile of the atmosphere within this model.
\citet{Lines2018} show that clouds form throughout most parts of HD~189733b's atmosphere, i.e. the local atmospheric temperature is below the condensation temperature of the Mg/Si/Fe/O/Ti cloud species considered.

In this paper, we utilize the vertical 1D ($T_{\rm gas}, p_{\rm gas}, K_{\rm zz}$)-profiles for the antistellar and the substellar points on the equator.
Figure~\ref{profiles_kzz_HD189733b} (left) shows that the 1D profiles appear remarkably similar throughout the whole pressure range covered by our computational domain, $p_{\rm gas}=10^2 - 10^{-5.5}\si{\bar}$.
Temperature differences of some $\SI{100}{\kelvin}$ occur in the uppermost region near the upper boundary of the 3D GCM model.
The eddy diffusion coefficient, $K_{\rm zz}$, varies between $10^8 \si{\centi\metre\squared\per\second}$ at $p_{\rm gas}=10^2\si{\bar}$ and  $10^{10} - 10^{11}\si{\centi\metre\squared\per\second}$ where $p_{\rm gas}<10^{-2}\si{\bar}$ (Fig.~\ref{profiles_kzz_HD189733b}, right).
Figure~\ref{Cloud_numdens_size_HD189733b_new} shows the cloud particle number density (left) and the mean cloud particles sizes (right).
Based on the 3D UM GCM, the cloud particle number density is rather homogeneous while all profiles show a strong accumulation of cloud particles due to gravitation settling at the inner boundary near $\SI{100}{\bar}$.
Within the 3D UM GCM, the cloud particles are of size $\SI{0.1}{\micro\metre}$ throughout the atmosphere.

\subsection{XUV spectra}
\label{SubSec_XUV}

\begin{figure*}
    \centering
    \includegraphics[width=\textwidth]{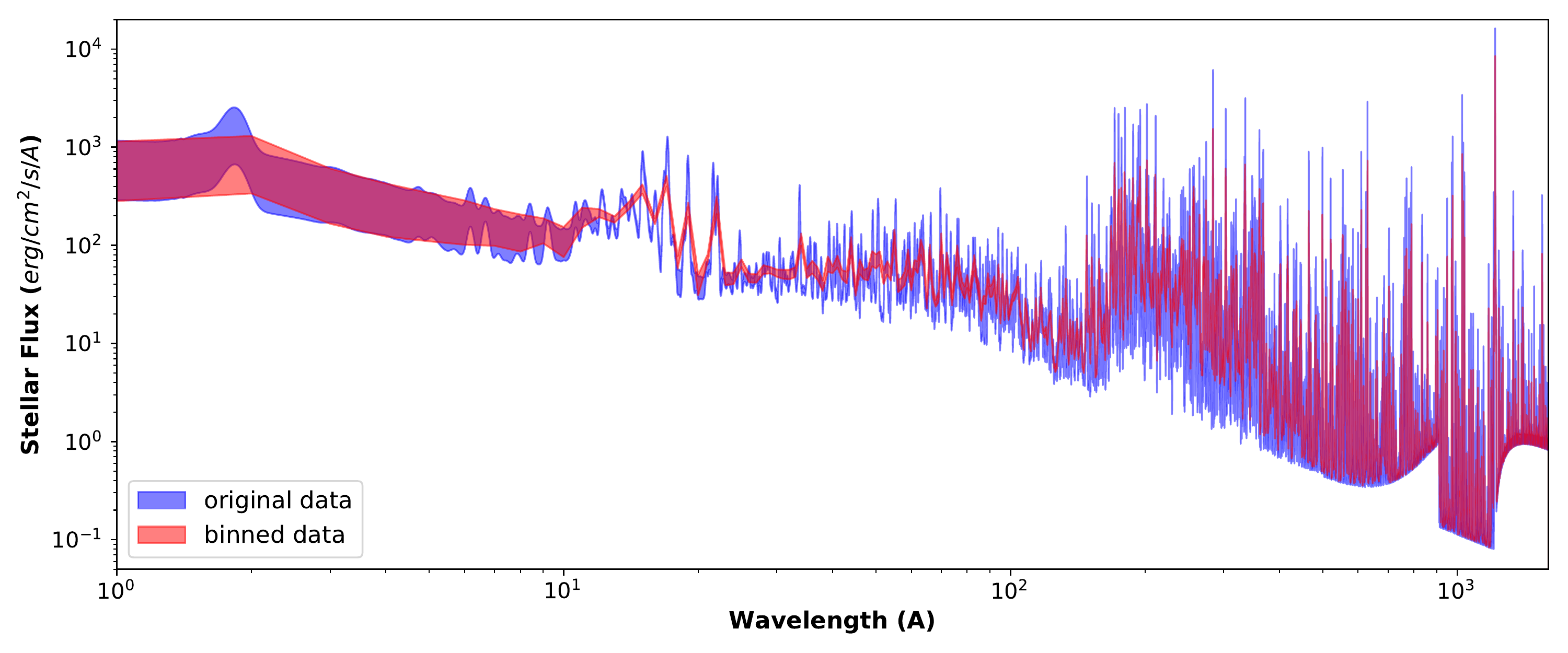}
    \caption{Original and binned XUV spectra of HD~189733 at the orbit of HD~189733b. Shown is the range in flux covered by the spectra from four different visits. The semi-synthetic spectra are derived from observations with \textit{HST}, \textit{XMM-Newton}, and \textit{Swift} \citep{Bourrier2020}.}
    \label{XUV_spec_original}
\end{figure*}

This work exploits XUV spectra of HD~189733 obtained within the framework of the MOVES collaboration \citep{Bourrier2020}.
UV (\textit{HST}) and X-ray (\textit{XMM-Newton/Swift}) observations were obtained in four contemporaneous epochs between July 2011 and November 2013.
After excluding flares, these observations provided measurements of the quiescent stellar emission in the soft X-ray (from the coronal region) and in a sample of FUV emission lines (from the chromosphere and transition region).
Most of the stellar EUV spectrum is not observable from Earth because of interstellar medium absorption.
The quiescent measurements were thus used to constrain a model of HD189733's atmosphere and to reconstruct its entire XUV spectrum in each epoch, using the Differentiel Emission Measure retrieval technique described in \citet{Louden2017}.
The Ly~$\alpha$ line of HD 189733 \citep[reconstructed independently following the procedure described in][]{Bourrier2017} which represents by itself half of the flux emitted in the entire EUV domain, has also been included in the synthetic spectra.
The final semi-synthetic spectra cover a wavelength range from 0.016 to $\SI{1600}{\angstrom}$ and therefore include not only the EUV but also the FUV flux \citep[these XUV spectra are available online as machine readable tables, see][]{Bourrier2020}.
The spread of the spectra is shown in Fig. \ref{XUV_spec_original}, along with their binned version (with a width of $\Delta \lambda = \SI{1}{\angstrom}$) used as input into ARGO.
The long-term monitoring of HD~189733 revealed variations in the spectral energy distribution of HD~189733 over these four epochs, with an increase in X-ray emission and a decrease in EUV emission, which could possibly trace a decrease in the chromospheric activity of the star while its corona became more active.

\begin{figure}
    \centering
    \includegraphics[width=\columnwidth]{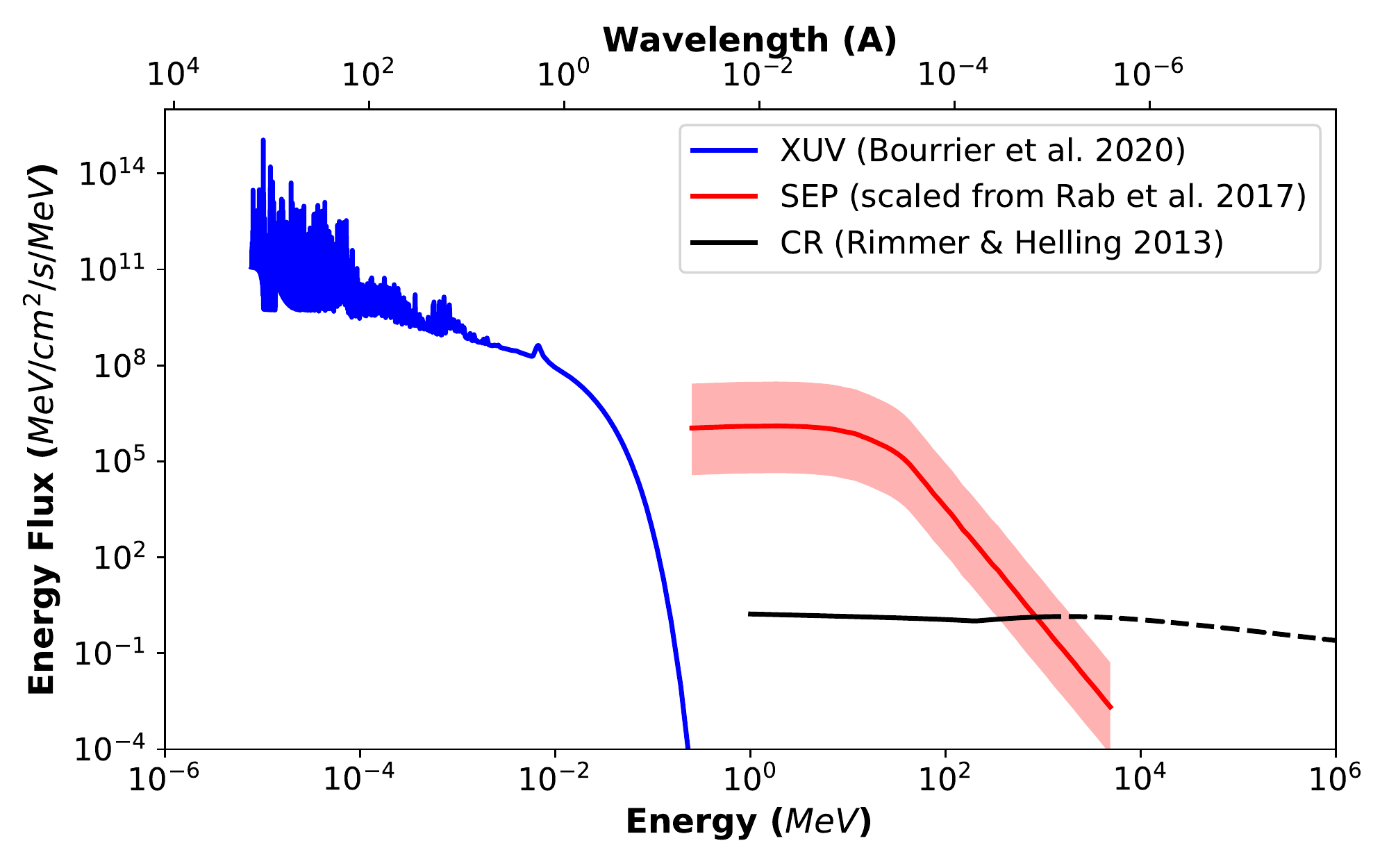}
    \caption{Comparison of high-energy spectra: XUV flux from HD~189733 \citep[blue line]{Bourrier2020}, stellar energetic particle spectrum from HD~189733 at the planet's orbit \citep[scaled from][red line]{Rab2017}, and low energy cosmic ray spectrum \citep[solid black line]{Rimmer2013a}. Cosmic rays with energies above $\SI{1}{\giga\electronvolt}$ are not included in the model (dashed black line). All values are scaled to the orbit of HD~189733b. Please note that these fluxes will most likely not occur at the same time and not on every part of the planet's surface. The SEP spectrum is only valid for a particle event associated with an X-ray flare while the XUV flux and the galactic cosmic rays are continuous energy sources.}
    \label{Fig_Energies}
\end{figure}
Figure~\ref{Fig_Energies} shows the XUV flux in comparison to the stellar energetic particles and the cosmic ray flux discussed in the following sections.

\subsection{Cosmic rays}
\label{SubSec_CR}

The effect of cosmic rays and their ionization rates of the atmosphere is implemented into \argo{}.
In our simulations, we include the effect of low energy cosmic rays (LECR) based on \citet{Rimmer2013a}.
The flux of cosmic-ray particles $j(E)$ is given by
\begin{equation}
  j(E) = \begin{cases}
    j(E_1) \left( \frac{p(E)}{p(E_1)} \right)^\gamma & \text{if $E > E_2$}\\
    j(E_1) \left( \frac{p(E_2)}{p(E_1)} \right)^\gamma \left( \frac{p(E)}{p(E_2)} \right)^\alpha & \text{if $E_\mathrm{cut} < E < E_2$}\\
    0 & \text{if $E < E_\mathrm{cut}$}
  \end{cases},
\end{equation}
with $p(E) = 1/c \sqrt{E^2 + 2 E E_0}$ and the proton rest energy $E_0 = 9.38 \times 10^8 \si{\eV}$, the constants $E_1 = 10^9 \si{\eV}$ and $E_2 = 2 \times 10^8 \si{\eV}$, and the flux $j(E_1) = 0.22 \si{\per\cm\squared\per\second\per\sterradian}$.
$E_\mathrm{cut} = 10^6 \si{\eV}$ is a low-energy cut-off.
Cosmic rays with energies lower than $E_\mathrm{cut}$ are unlikely to travel further from their source than $\sim \SI{1}{\parsec}$ \citep{Rimmer2012,Rimmer2013a}.

The effect of these cosmic rays on the atmosphere is parameterized by the ionization rate $\zeta$ which depends on the column density $N_\mathrm{col} [\si{\per\centi\metre\squared}]$ of the gas in the atmosphere. $\zeta$ is well fitted by:
\begin{equation}
  \zeta (N_\mathrm{col}) = \zeta_0 \times \begin{cases}
    480 & \text{if $N_\mathrm{col} < N_1$}\\
    1 + (N_0/N_\mathrm{col})^{1.2} & \text{if $N_1 < N_\mathrm{col} < N_2$}\\
    \exp{(-k N_\mathrm{col})} & \text{if $N_\mathrm{col} < N_2$}
  \end{cases},
\end{equation}
with the standard ionization rate in the dense interstellar medium $\zeta_0 = 10^{-17} \si{\per\second}$, the column densities $N_0 = 7.85 \times 10^{21} \si{\per\centi\metre\squared}$, $N_1 = 4.6 \times 10^{19} \si{\per\centi\metre\squared}$, $N_2 = 5.0 \times 10^{23} \si{\per\centi\metre\squared}$, and $k = 1.7 \times 10^{-26} \si{\centi\metre\squared}$.
The ionization rate is shown in Fig.~\ref{HD_proton_flux} (lower panel, dotted black line).

\subsection{Stellar Energetic Particles}
\label{SubSec_SEP}

\begin{figure}
    \centering
    \includegraphics[width=\columnwidth,page=1]{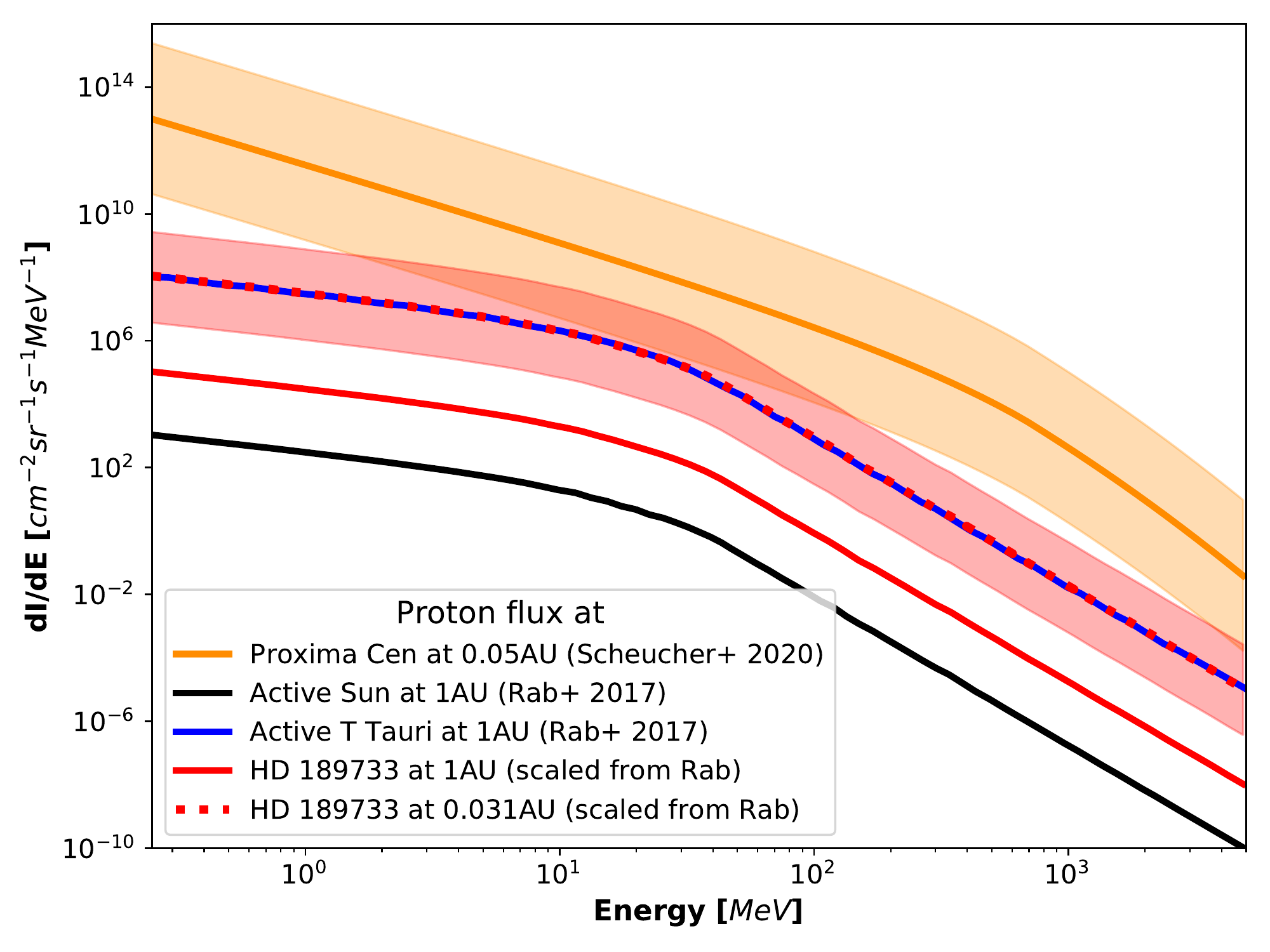}\\
    \includegraphics[width=\columnwidth,page=2]{HD_proton_flux.pdf}
    \caption{\textit{Top:} Stellar energetic particle (SEP) spectrum from HD~189733 at $\SI{1}{\astronomicalunit}$ (red solid) and at the planet's orbit ($\SI{0.031}{\astronomicalunit}$, red dashed) are shown alongside the spectrum for an active Sun (black) and an active T Tauri star (blue, both at $\SI{1}{\astronomicalunit}$) from \citet{Rab2017}. These were used to estimate the spectrum for HD~189733. For comparison the SEP spectrum at Proxima Centauri b's orbit is shown \citep{Scheucher2020}. The shaded areas indicate the error range based on the peak size distribution function in \citet{Herbst2019}.
    \textit{Bottom:} The resulting ionization rates against the atmospheric column density are shown for the different spectra. For comparison the low energy cosmic rays (LECR) ionization rate from \citet{Rimmer2013a} is shown (black dotted).}
    \label{HD_proton_flux}
\end{figure}

Stellar energetic particle (SEP) events are caused by solar flares and coronal mass ejections.
These events are accompanied by flares in the X-ray and UV emission and the peak size distribution of these flares follows a power law \citep{Hudson1978,Belov2005a}.
\citet{Herbst2019} present a new peak size distribution that allows an estimate of the proton flux associated with flares on G-, K-, and M-dwarfs:
\begin{equation}
    I_\mathrm{p} (E > \SI{10}{\mega\electronvolt}) = (a I_\mathrm{x} + b \exp{(-0.001 I_\mathrm{x})})^\gamma,
    \label{Eq_PSD_Herbst}
\end{equation}
with $a = 1.22 \times 10^5 (\pm 7.17\times 10^4)$, $b = 3.05 \pm 1.79$, and $\gamma = 1.72 \pm 0.397$.

\citet{Pillitteri2010} provides observations with \textit{XMM-Newton} of an X-ray flare on HD~189733.
In 2009, they measured an increase in the $0.3 - \SI{8}{\kilo\electronvolt}$ band due to a flare of $1.3 \times 10^{-13} \si{\erg\per\second\per\centi\metre\squared}$.
At the planet's orbit this corresponds to an X-ray flare intensity of $I_\mathrm{x} = \SI{2.52}{\watt\per\metre\squared}$.
Using Eq.~(\ref{Eq_PSD_Herbst}) this gives a proton flux\footnote{Please note that Eq.~(\ref{Eq_PSD_Herbst}) is only valid for fluxes at a distance of $\SI{1}{\astronomicalunit}$ to the star.
For HD~189733b, $I_\mathrm{x}$ has to be scaled with $r^2$ to $\SI{1}{\astronomicalunit}$ and then the resulting $I_\mathrm{p}$ back to the planet's distance to its host star.}
of $I_\mathrm{p} = 1.55 (+ 36.4 - 1.50) \times 10^7 \si{\per\centi\metre\squared\per\sterradian\per\second}$.

We derive the spectrum of the SEP event from the integrated proton flux by using the fitted spectrum of a solar SEP event from 2003 \citep{Mewaldt2005}.
\citet{Rab2017} used this spectrum to estimate the SEP spectrum of an active T Tauri star by scaling the solar spectrum by a factor of $10^5$.
The integrated proton flux of the solar event spectrum is $I_\mathrm{p} (E > \SI{10}{\mega\electronvolt}) = 151 \si{\per\centi\metre\squared\per\sterradian\per\second}$.
Scaling the solar SEP spectrum such that the integrated flux corresponds to the value we calculated for HD~189733 above results in the spectrum shown in Fig.~\ref{HD_proton_flux} (top panel).
The dotted red line represents the SEP event at the orbit of HD~189733b and coincides with the spectrum for the active T Tauri star at a distance from the star of $\SI{1}{\astronomicalunit}$.
This allows us to use both the spectrum and the parametrized ionization rate (below) for the active T Tauri star from \citet{Rab2017}.
Please note, that this is only the case because HD~189733b is much closer to its host star than $\SI{1}{\astronomicalunit}$.
At $\SI{1}{\astronomicalunit}$, the proton flux from HD~189733 is much smaller than the flux from the active T Tauri star (solid red line).
\citet{Bourrier2020} also observed an X-ray flare of HD~189733 which has a slightly higher luminosity than the one observed by \citep{Pillitteri2010}, with an X-ray flare intensity at the planet's orbit of $I_\mathrm{x} = \SI{7.4}{\watt\per\metre\squared}$.

Once we have an SEP spectrum for HD~189733 we can calculate the ionization rate $\zeta_\mathrm{SEP}$ as a function of the total hydrogen column density $N_{<\ce{H}>} = N_{\ce{H}} + 2N_{\ce{H2}} + 2N_{\ce{H2O}} + N_{\ce{H^+}}$ \citep[their Eq.~1 \& 2]{Rab2017}:
\begin{multline}
    \zeta_\mathrm{SEP}(N_{<\ce{H}>}) = \zeta_\mathrm{SEP,0}(N_{<\ce{H}>}) \\ \times
        \begin{cases}
            1 & \text{if $N_{<\ce{H}>} \le N_\mathrm{E}$} \\
            \mathrm{exp}\left[ - \left( \frac{N_{<\ce{H}>}}{N_\mathrm{E}} -1 \right) \right] & \text{if $N_{<\ce{H}>} > N_\mathrm{E}$}
    \end{cases},
\end{multline}
with
\begin{equation}
    \zeta_\mathrm{SEP,0}(N_{<\ce{H}>}) = \left[
        \frac{1}{\zeta_\mathrm{L} \left(
            \frac{N_{<\ce{H}>}}{10^{20} \si{\per\centi\metre\squared}}
            \right)^a} +
        \frac{1}{\zeta_\mathrm{H} \left(
            \frac{N_{<\ce{H}>}}{10^{20} \si{\per\centi\metre\squared}}
            \right)^b}
        \right]^{-1},
\end{equation}
$N_\mathrm{E} = 2.5 \times 10^{25} \si{\per\centi\metre\squared}$, $a = -0.61$, $b = -2.61$, $\zeta_\mathrm{L} = 1.06 \times 10^l \si{\per\second}$, and $\zeta_\mathrm{H} = 8.34 \times 10^h \si{\per\second}$.
The exponents $l$ and $h$ depend on the peak proton flux at the top of the atmosphere and are listed in Table~\ref{Tab_SEP_ion}.
The resulting ionization rates are shown in Fig.~\ref{HD_proton_flux} (lower panel) together with the ionization rate by the low energy cosmic rays as presented in Section~\ref{SubSec_CR}.

\begin{table}
    \caption{Fitting parameters for the stellar energetic particle (SEP) ionization rates.}
	\begin{tabular}{lcc}
		\noalign{\smallskip}
		\hline
		\noalign{\smallskip}
		Name & $l$ & $h$\\
		\noalign{\smallskip}
		\hline \hline
		\noalign{\smallskip}
		Active Sun at $\SI{1}{\astronomicalunit}$ $^a$ & -12 & -7 \\
		Active T Tauri at $\SI{1}{\astronomicalunit}$ $^a$ & -7 & -2 \\
		HD~189733 at $\SI{1}{\astronomicalunit}$ & -7.1 & -2.1 \\
		HD~189733 at $\SI{0.031}{\astronomicalunit}$ & -10.0 & -5.0 \\
		\noalign{\smallskip}
		\hline
	\end{tabular}\\
	 $^a$ \citet{Rab2017}
	\label{Tab_SEP_ion}
\end{table}

\section{Atmospheric C/H/N/O composition under the effect of XUV, SEP, and CR on HD~189733\texorpdfstring{\MakeLowercase{b}}{b}}
\label{Sec_Results}

\begin{figure*}
    \centering
    \includegraphics[page=1,width=\textwidth]{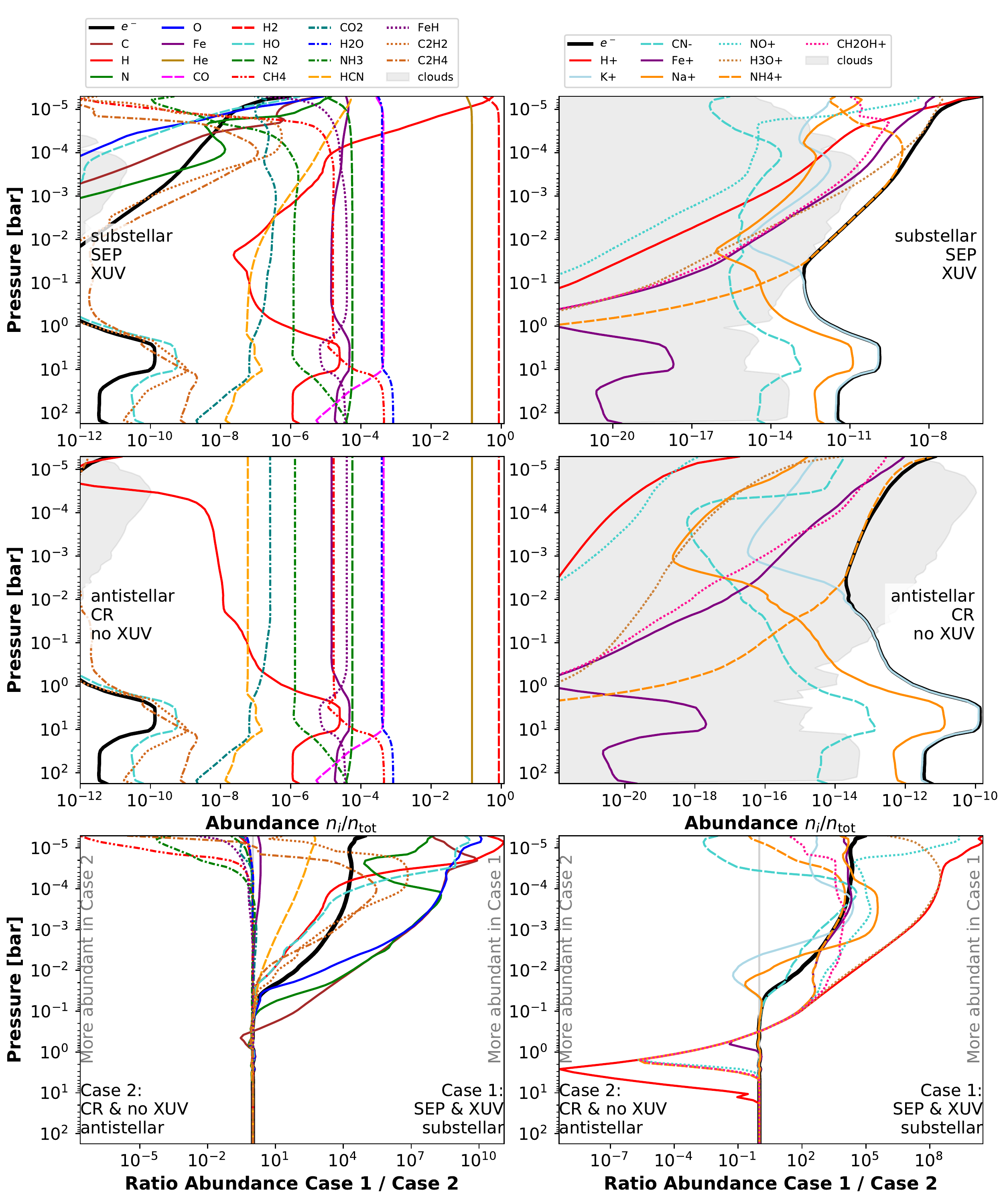}
    \caption{Atmospheric composition of HD~189733b at the day- and nightside profiles: \textit{Top:} substellar point with SEP and XUV radiation (case 1). \textit{Middle:} antistellar point with cosmic rays but no XUV radiation (case 2).
    Relative abundances ($n_i/n_\mathrm{tot}$) of most abundant neutral molecules (\textit{left}) and most abundant ions (\textit{right}) are shown.
    Thick, black line is the abundance of electrons and gives an estimate on the degree of ionization.
    Grey area indicates the abundance of cloud particles.
    \textit{Bottom:} Ratio of abundances of most abundant neutral molecules (\textit{left}) and ions (\textit{right}) comparing the two cases shown in the panels above. Please note that the abundances (x axis) change from panel to panel.}
    \label{HD189733b_abundances}
\end{figure*}

In this section, we present the atmospheric H/C/N/O gas-phase abundances of HD~189733b for two different profiles: the substellar and antistellar point, i.e. the day- and nightside, respectively. We will examine the effects of the XUV flux, stellar energetic particles, and cosmic rays on the abundance of neutral molecules and ions.

First, we investigate the effect that the observed changing XUV and FUV radiation (in the following we assume XUV to include FUV raditaion) of the host star HD~189733 \citep{Bourrier2020} has on its planet HD~189733b.
For this, we study the influence of the four different XUV spectra (see Sec. \ref{SubSec_XUV}) on the chemical composition of  planetary atmosphere.
We concentrate our study on H/C/N/O binding species, metal atoms and their singly ionized states.
The variations in the  quiescent XUV flux of the host star HD~189733 are too small to significantly change the planet's gas composition.
Hence, any variation of the gas-phase abundances must originate from other processes such as flares leading to SEP events.
In the following, we will therefore present results for one XUV spectrum only, which is from an observation in September 2011 \citep[Visit B in][]{Bourrier2020}.

For the initial element abundances in the atmosphere we assume solar element abundances \citep{Asplund2009}.
As clouds form throughout the atmosphere, we account for the  depletion of \ce{Mg}, \ce{Si}, and \ce{Ti} due to consumption by cloud particle formation (\ce{TiO2}[s], \ce{SiO}[s], \ce{SiO2}[s], \ce{MgSiO3}[s], and \ce{Mg2SiO4}[s]).

\subsection{Asymmetric irradiation of day- and nightside}
\label{subsec_assym}

The dayside is irradiated by the stellar XUV radiation and the SEP flux.
We do not take into account the galactic cosmic rays (CR) on the dayside since the particle flux is dominated by SEPs for most of the spectrum.
Furthermore, CRs are more likely to be shielded by the star's astrosphere on the dayside rather than on the nightside \citep{Rimmer2016}.
The nightside in turn receives only the influx of CRs.
We note that the influx of SEPs into the atmosphere is a function of time and will not be present at all times.
Figure~\ref{HD189733b_abundances} shows the results of our simulations for both the substellar (top row, case 1) and the antistellar point (middle row, case 2) including the SEP and XUV flux on the dayside and the CR flux on the nightside.
The bottom row shows the ratios of the abundances between the substellar and antistellar abundances.
The abundances of the most important neutral elements are shown in the left column, the abundances of the most important ions in the right column.

The asymmetric irradiation of the planet's day- and nightside leads to strong differences between the chemical abundances of these profiles.
While on the nightside the low energy input from cosmic rays leads to only marginal differences compared to the equilibrium case, the high SEP flux and XUV radiation on the dayside enhance atomic species (H, C, N, and O) and increase the abundance of HCN and several hydrocarbons such as \ce{C2H2} and \ce{C2H4}.

We note that the atmospheres of close-in, tidally locked gas giants are characterised by strong equatorial jets \citep[e.g.][]{Cooper2006,Drummond_2018b,Carone2020}.
Such jets lead to very similar day/night vertical temperature distributions (see Fig.~\ref{profiles_kzz_HD189733b}).
\citet{Drummond_2020} showed that a strong equatorial jet with zonal-mean velocities of $\sim \SI{6}{\kilo\metre\per\second}$ may result in kinetic quenching of the methane abundances above $\sim \SI{1}{\bar}$ for all three wind components (zonal, meridional, and vertical).
Unlike in their previous work \citep{Drummond2018b} where they applied a simple chemical relaxation scheme by \citet{Cooper2006}, \citet{Drummond_2020} used a more accurate coupled chemical kinetics scheme similar to what we use in this study.
However, they used a reduced chemical network of \citet{Venot2019}, containing only 30 species and 181 reactions.

\citet{Agundez2014} used a 'pseudo-2D' model, moving a 1D column with vertical mixing along the equator to imitate a solid-body rotation of the atmosphere, including photochemistry.
They used a chemical network with 105 species developed by \citet{Venot2012a} and found that horizontal quenching enhances the abundance of \ce{HCN} on the nightside by transporting photochemically produced \ce{HCN} from the day- to the nightside.
A first order comparison of the destruction time scale of \ce{HCN}, \ce{CH2O}, and \ce{C2H4} to the transport time-scale from the sub- to the antistellar point indicates that the destruction time-scales are orders of magnitudes shorter than the transport time.
At $\SI{6}{\kilo\metre\per\second}$, gas will be transported from the sub- to the antistellar point in approximately $\SI{12}{h}$.
In comparison to that, the destruction rates of \ce{HCN}, \ce{CH2O}, and \ce{C2H4} are much faster, meaning it would take only 16, 0.3, and $\SI{5}{min}$, respectively, to destroy the dayside abundances of these molecules.
However, \citet{Moses2011} showed that changes in the abundance of \ce{NH3} also impact the abundance of other nitrogen-bearing species even if they would otherwise not be quenched.
Thus \ce{HCN} remains in a pseudo-equilibrium with \ce{NH3} and \ce{CH4} and the potential horizontal quenching of these species has to be taken into account as well.

\subsection{The atmospheric electron production rate and fingerprint ions}
\label{ss:els}

\begin{figure*}
    \centering
    \includegraphics[width=0.7\textwidth]{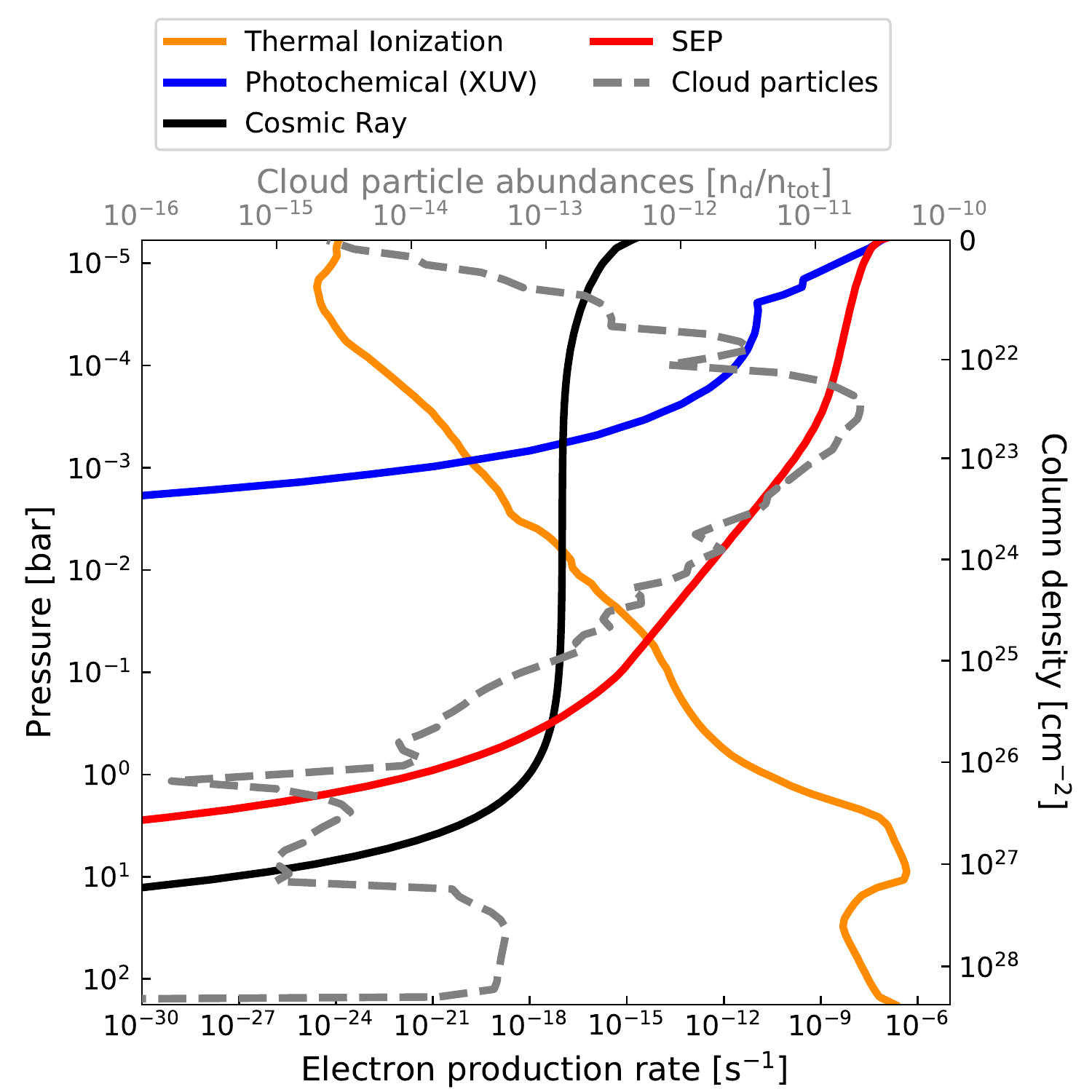}
    \caption{Overview of electron production rates as a proxy for the ionization rates of the individual ionization sources plotted against the atmospheric pressure and the atmospheric column density. Cloud particle abundances (grey) are added for comparison as the position of clouds and electrons are important for the presence of lightning. The temperature profile and cloud particles used for these simulations are for the substellar point. Note that the column density at the top of the atmosphere is by definition 0 since we do not have data on the layers further up.}
    \label{Fig_ioniziation_compare}
\end{figure*}

The ionization of the atmospheres of exoplanets and the electron production rates by XUV radiation and high-energy particles are linked to various high-energy processes.
One example is the production of auroral emission that might be observable with future telescopes and help us understand the magnetic field of the planet \citep{Burkhart2017,Luger2017,Vedantham2020}.
Furthermore, ion production in the atmosphere will also ionize cloud particles which is necessary for the occurrence of lightning:
the larger a cloud particle, the higher the charge it acquires and the faster it will gravitationally settle within its host cloud.
Such particle settling leads to the build-up of a local electric field in the cloud, and if the field is large enough, a lightning strike will occur \citep{Helling2011a,Helling2011,Helling2019}.
In an Earth-like atmosphere, a lightning strike produces an approximately $\SI{30000}{\kelvin}$ hot plasma channel which only exists for a few seconds but has a strong influence on the chemistry in the surrounding atmosphere \citep{An2019}.
The detailed effect of lightning on the atmospheric chemistry will be the focus of future studies.

To investigate the ionization environment of HD~189733b's atmosphere, we calculate the electron production rates for all included ionization sources (XUV, SEP, and CR radiation as well as thermal ionization).
Figure~\ref{Fig_ioniziation_compare} shows a comparison of these ionization rates.
In addition, the cloud particle density is shown to get an idea of the cloud position in comparison to the charge positions.

The lower atmosphere up to $10 - \SI{100}{\milli\bar}$ is in all cases dominated by thermal ionization.
Above $10 - \SI{100}{\milli\bar}$, ionization by external sources becomes more important.
The electron production rates by XUV radiation, SEP, and CR can be explained by the shape of the individual spectra (Fig.~\ref{Fig_Energies}):
the XUV flux is very high but at the lower end of the energy spectrum.
Most of the energy is deposited in the uppermost layers of the atmosphere, and comparatively few photons will be able to ionize deeper layers.
The SEP flux is a bit lower but at much higher energies and is therefore able to penetrate deeper into the atmosphere.
The CR flux is much lower, leading to a correspondingly lower electron production rate compared to that of other radiation sources.
Around $\SI{1}{\giga\electronvolt}$, the CR flux is higher than the SEP flux, meaning that CRs can reach even deeper levels of the atmosphere.
However, in these layers, the thermal ionization rate is much higher than that of the CRs.
Expanding the parametrization of the CR ionization rate to high energy CR ($>\SI{1}{\giga\electronvolt}$) would only shift the drop off of the CR electron production rate to higher pressures but would not increase the electron production rate itself.

The most important electron donors in the deep atmosphere ($p \gtrsim 10^{-2} - 10^{-1} \si{\bar}$) are potassium and sodium (Fig.~\ref{HD189733b_abundances}, right top and center panel).
The most abundant ion produced by both CR and SEP is ammonium \ce{NH4^+}.
In addition, the high ionization rates associated with the SEP flux and to a lesser extent with XUV radiation, produces a larger variety of ions: \ce{H3O^+} and \ce{H^+}, as well as smaller amounts of \ce{Fe^+} and \ce{CH2OH^+}.
Therefore, we identify \ce{NH4^+} and \ce{H3O^+} as the fingerprint ions for ionization by SEP and \ce{NH4^+} by CR.

\subsection{The effect of high-energy radiation on the production of organic, prebiotic molecules}
\label{ss:prebio}

Ultimately, we are interested in the effect of the different types of high-energy radiation on the organic and prebiotic chemistry.
The famous experiment by Stanley Miller and Harold Urey suggests that lightning is an important source of organic molecules, at least in a reducing (i.e. hydrogen rich) atmosphere \citep{Miller1953}.
Cloud-forming exoplanets such as HD~189733b are expected to produce lightning due to their very dynamic and cloudy atmosphere \citep{Lee2016,Lines2018}.
Before we can study the effect of lightning on the atmospheric chemistry and identify observable signatures, we need to understand the effect of other types of high-energy radiation on potential lightning signatures in the chemical abundances.

\begin{figure}
    \centering
    \includegraphics[width=\columnwidth]{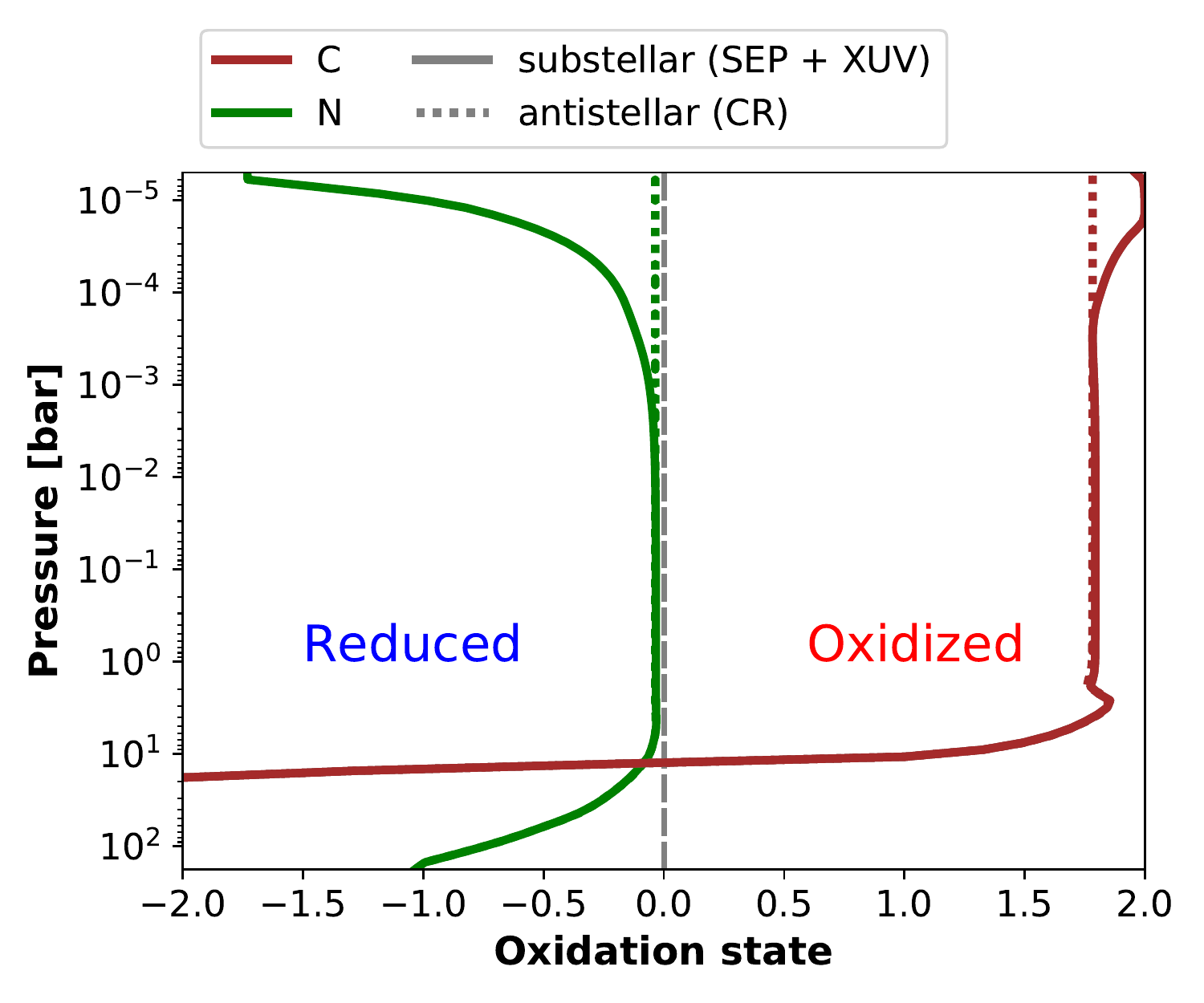}
    \caption{Average oxidation state of nitrogen (\textit{green}) and carbon (\textit{brown}) atoms in the atmosphere of HD~189733b at the substellar (\textit{solid}, SEP + XUV) and antistellar point (\textit{dotted}, CR). Cosmic rays alone do not change the oxidation state of carbon and nitrogen, i.e. the antistellar profiles are the same as the equilibrium case. The dashed grey line indicates a neutral oxidation state. Negative values mean the nitrogen and carbon are reduced, positive values indicate a oxidized state.}
    \label{Nitro_redox}
\end{figure}

The average oxidation state of the carbon and nitrogen atoms in the C and N bearing molecules provides an overview over the processes in the atmosphere and their effect on these molecules.
The oxidation state of an atom in a compound indicates the charge it would have if all bonds in the compound were ionic.
One example is the oxidation state of the carbon atom in different neutral carbon molecules:
\ce{CH4} is the most reduced form of carbon with an oxidation state of -4, i.e. the C atom is slightly more electronegative and attracts the electrons from each of the four H atoms.
\ce{CO2} in comparison is the most oxidized form of carbon with an oxidation state of +4 as the O atoms have a higher electronegativity than C.
As oxygen and hydrogen are, respectively, the most and least electronegative elements of the most abundant elements (the others being carbon and nitrogen, while helium is a noble gas and has therefore no defined electronegativity), the oxidation state of a gas can be approximated by the H/O ratio with H having an oxidation state of +1, i.e. reducing the C or N atom, and oxygen an oxidation state of -2.

Figure~\ref{Nitro_redox} shows the oxidation states of the C and N atoms at HD~189733b's sub- and antistellar point.
The most abundant nitrogen species is \ce{N2} which is by definition neither oxidized nor reduced (oxidation state of 0).
Therefore, the average oxidation state of the nitrogen atoms is close to zero.
The second most abundant nitrogen molecule in most of the atmosphere is \ce{NH3}, where the nitrogen is reduced (-3).
The SEP strongly enhance the abundance of \ce{HCN} in the upper atmosphere (Fig.~\ref{HD189733b_abundances}), further reducing the nitrogen.
Importantly, life as we know it requires reduced N for the production of biomolecules such as amino acids and nucleobases. High-energy radiation may therefore make atmospheric N more bioavailable.

The effect on the average oxidation state of carbon is very different.
Except for the very deep parts of the atmosphere of HD~189733b ($p>\SI{10}{\bar}$), where methane is the most abundant carbon molecule and therefore reducing the atmospheric carbon, most of the carbon is bound in \ce{CO} which has an oxidation state of +2.
The enhanced abundance of \ce{HCN} that further reduces the atmospheric nitrogen when adding SEP radiation, leads to an additional oxidation of carbon in the upper atmosphere. In biomolecules, the oxidation state of organic carbon is variable with an average around 0.
Both oxidized and reduced carbon can act as metabolites for autotrophic life and is therefore readily bioavailable.

\begin{figure*}
    \centering
    \includegraphics[width=\textwidth]{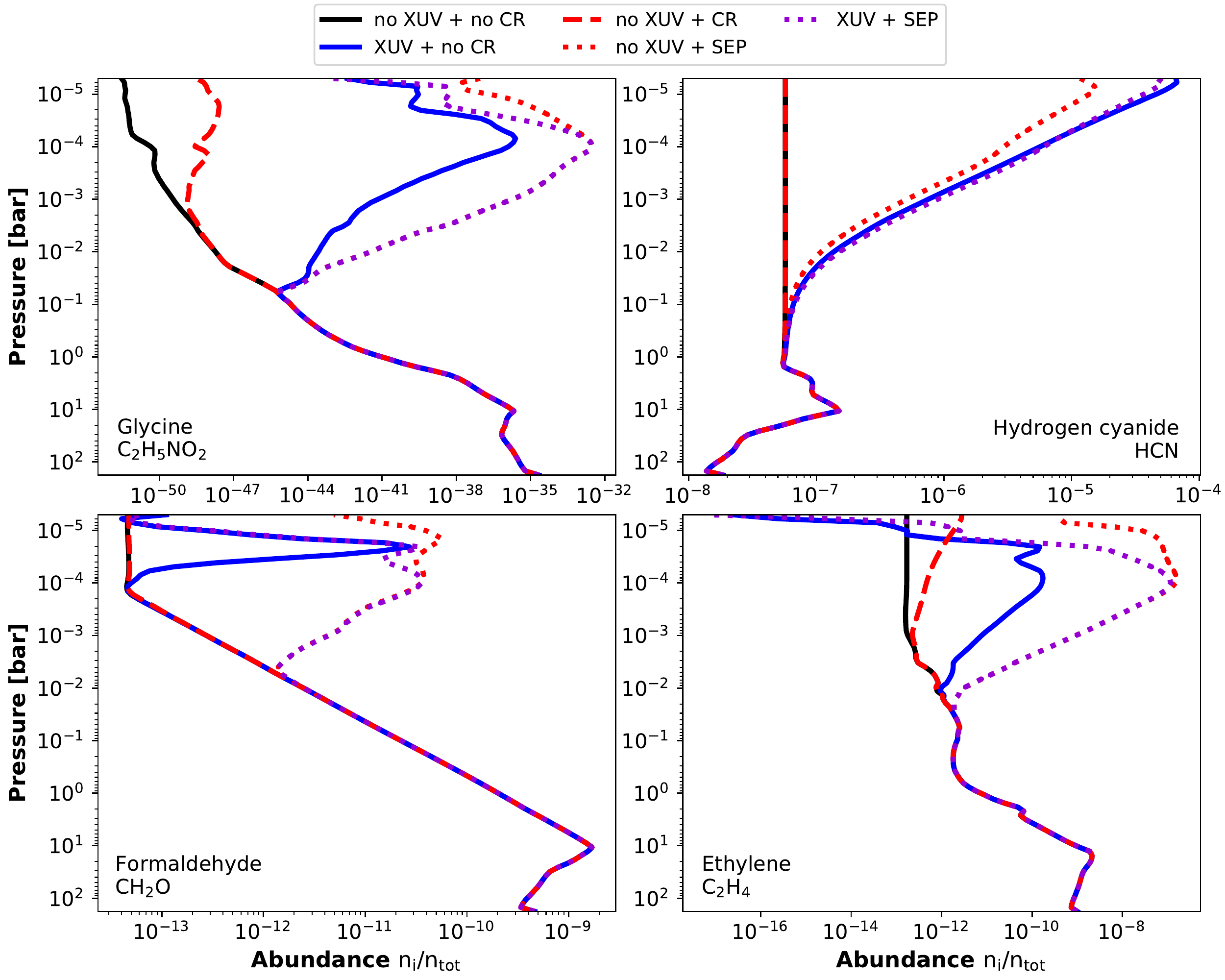}
    \caption{Abundance of glycine (\textit{top left}), \ce{HCN} (\textit{top right}), formaldehyde (\ce{CH2O}, \textit{bottom left}), and ethylene (\ce{C2H4}, \textit{bottom right}) in the atmosphere of HD~189733b at the substellar point with different combinations of the ionization sources XUV radiation, SEP, and CR. We note that a lightning induced HCN increase would be best observable on the nightside as we do not see any enhancement by other ionization sources. Note the different ranges on the x axis.}
    \label{HD189733b_glycine}
\end{figure*}

\begin{figure*}
    \centering
    \includegraphics[page=1,width=\textwidth]{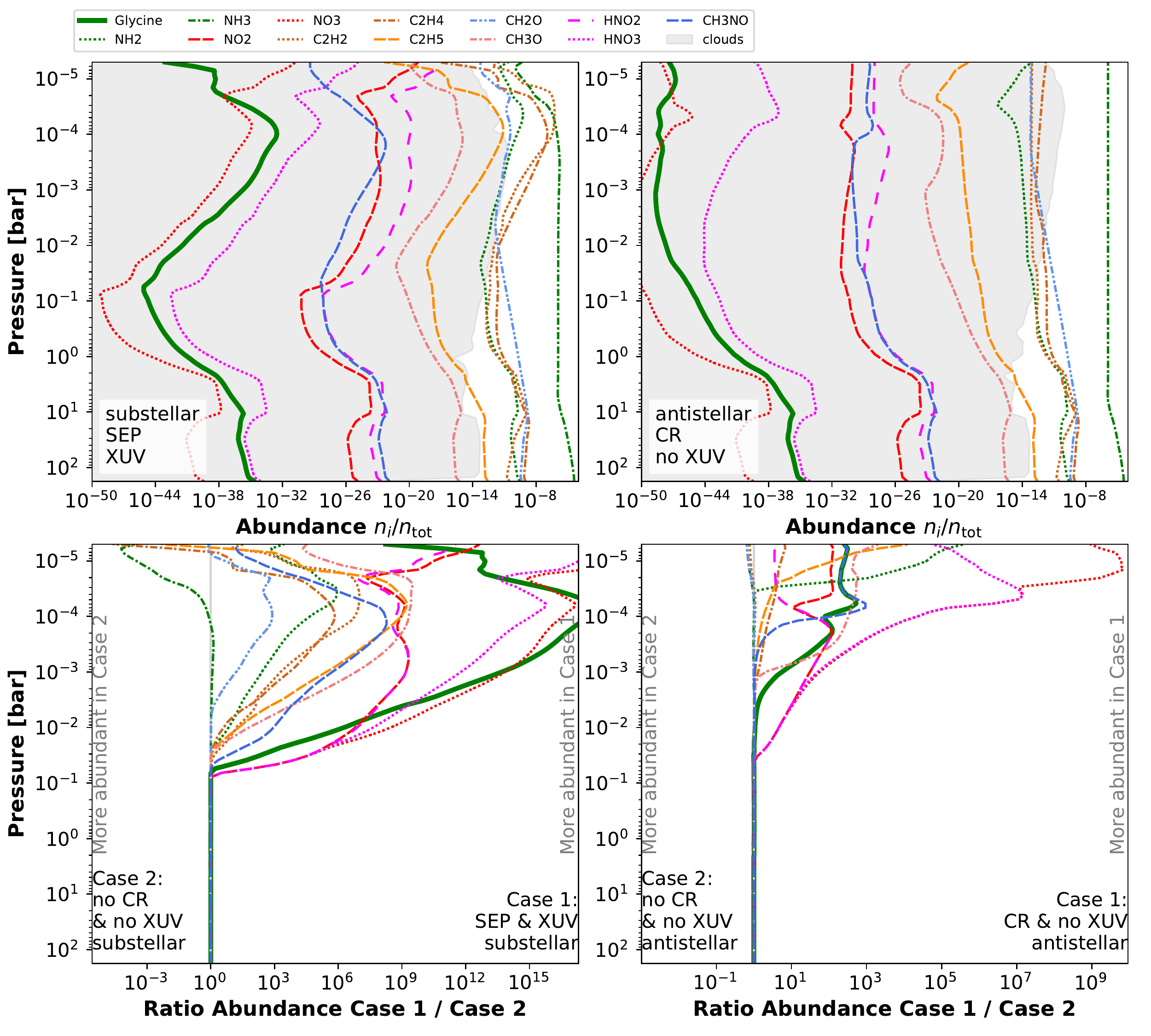}
    \caption{Abundances of molecules important for the formation of glycine (solid green) in the atmosphere of HD~189733b. \textit{Top left:} substellar point with SEP and XUV radiation (case 1). \textit{Top right:} antistellar point with cosmic rays but no XUV radiation (case 2). Grey shaded areas indicate the presence of clouds.
    \textit{Bottom:} Ratio of abundances shown in the panels above to the equilibrium case.}
    \label{HD189733b_abundances_lightning}
\end{figure*}

The species that dominates the change in the oxidation state of carbon and nitrogen is \ce{HCN} which contains a CN bond that is also found in the amino acid glycine and in other biomolecules.
Figure~\ref{HD189733b_glycine} shows the abundance of glycine (\ce{C2H5NO2}, top left) and \ce{HCN} (top right) with different combinations of external ionization sources XUV, CR, and SEP.
The results in Fig.~\ref{HD189733b_glycine} are shown just for the substellar point.
The antistellar profiles are not shown due to being very similar to the substellar profiles (see Fig.~\ref{profiles_kzz_HD189733b}).
The \stand{} network includes the complete gas-phase production reactions for glycine which was one of the amino acids produced in the original Miller--Urey experiment \citep{Miller1953}.

\begin{figure*}
    \centering
    \includegraphics[width=\textwidth]{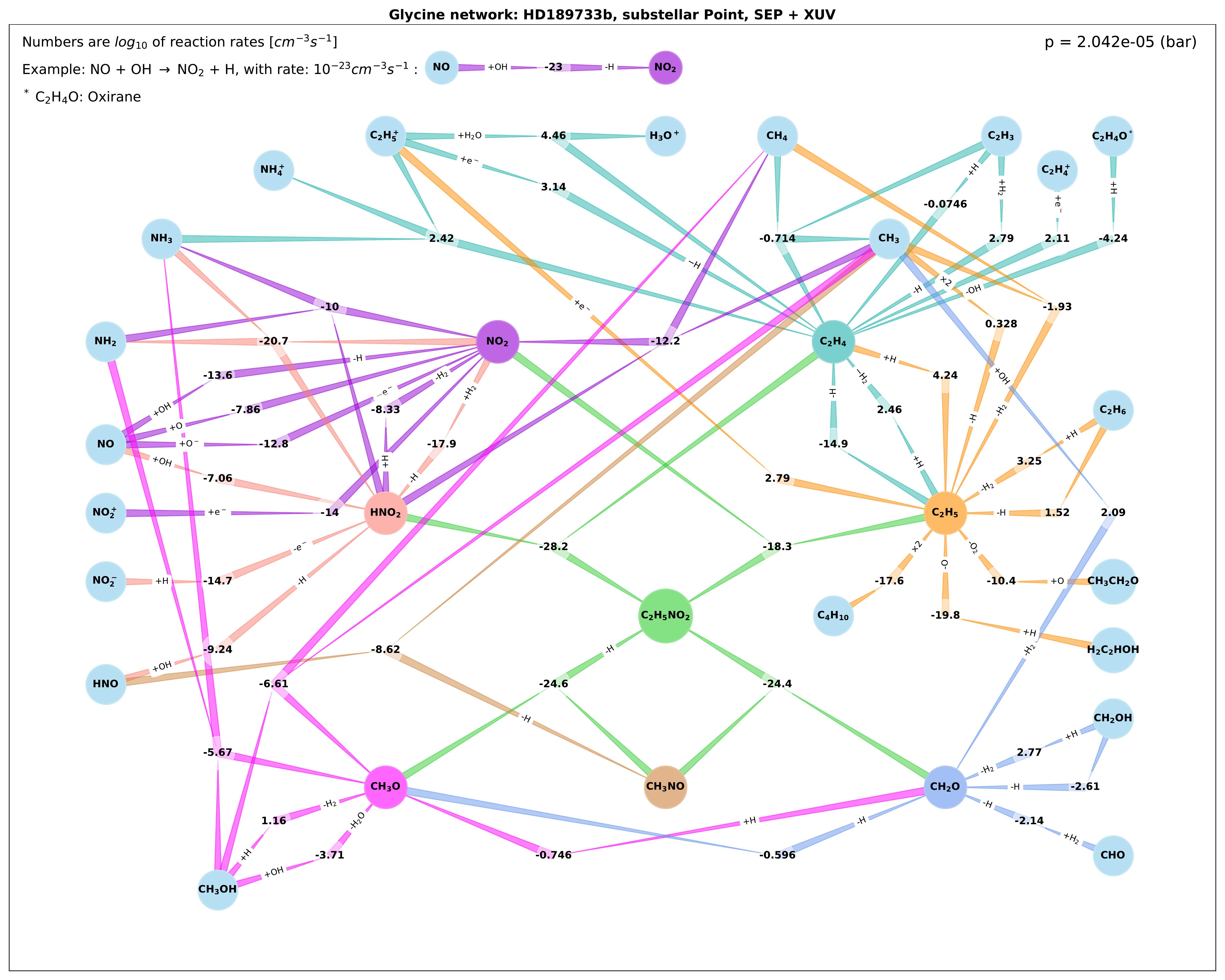}
    \caption{Production pathways of glycine (\ce{C2H5NO2}, green) and its precursors. Numbers are reaction rates ($\si{\per\centi\metre\cubed\per\second}$, $\log_{10}$) for forward reactions (in the direction towards the sharp end of the lines, mainly from the outer regions towards the centre of the network) producing the molecule in the same colour as the arrows and potential side products. Reaction rates for reverse reactions are not shown. Small, abundant reactants and products (e.g. \ce{OH}) are written along the reaction path to decrease number of paths; '+': molecule is added to reaction, '-': molecule is additional product of reaction, '$\times 2$': 2 \ce{CH3} molecules needed for production of \ce{C2H5}. Examples: \ce{NH2} and \ce{HNO2} (pink) produce \ce{NH3} and \ce{NO2} (purple) at a rate of $10^{-10}\si{\per\centi\metre\cubed\per\second}$; \ce{NO} and \ce{OH} produce \ce{NO2} (purple) and \ce{H} at a rate of $10^{-13.6}\si{\per\centi\metre\cubed\per\second}$.}
    \label{Network_glycine}
\end{figure*}

\begin{table}
    \caption{Pathways of glycine production at $p = 2\times10^{-5}\si{\bar}$ in the atmosphere of HD~189733b at the substellar and antistellar point, as well as for the equilibrium case. Colors correspond to path diagram (Fig.~\ref{Network_glycine}), bold numbers indicate most important pathways.}
	\begin{tabular}{lccc}
		\noalign{\smallskip}
		\hline
		\noalign{\smallskip}
		 & \multicolumn{3}{c}{log( Rate [$\si{\per\centi\metre\cubed\per\second}$])} \\
		\noalign{\smallskip}
		\hline
		\noalign{\smallskip}
		\textbf{Reaction} & Equ. & CR & SEP+XUV \\
		\noalign{\smallskip}
		\hline \hline
		\noalign{\smallskip}
		\ce{C2H5^+ + H2O -> \textcolor{SeaGreen}{\ce{C2H4}} + H3O^+} & -47.1 & \textbf{-3.47} & \textbf{4.46} \\
        \ce{C2H5^+ + NH3 -> \textcolor{SeaGreen}{\ce{C2H4}} + NH4^+} & -48.1 & -4.42 & 2.42 \\
        \ce{C2H5^+ + e^- -> \textcolor{SeaGreen}{\ce{C2H4}} + H}     & -55.1 & -9.22 & 3.14 \\
        \ce{C2H3 + H2    -> \textcolor{SeaGreen}{\ce{C2H4}} + H}     & \textbf{-8.32} & -6.73 & 2.79 \\
        \noalign{\smallskip}
        \hline
        \noalign{\smallskip}
        \ce{\textcolor{SeaGreen}{\ce{C2H4}} + H -> \textcolor{YellowOrange}{\ce{C2H5}}} & \textbf{-7.12} & \textbf{-9.29} & \textbf{4.24} \\
        \noalign{\smallskip}
        \hline \hline
        \noalign{\smallskip}
        \ce{NO + OH -> \textcolor{Lavender}{\ce{HNO2}}}      & \textbf{-21.7} & \textbf{-20.7} & -7.06 \\
        \noalign{\smallskip}
        \hline
        \noalign{\smallskip}
        \ce{NO + O -> \textcolor{Purple}{\ce{NO2}}}          & -31.4 & -21.6 & \textbf{-7.86}  \\
        \ce{\textcolor{Lavender}{\ce{HNO2}} + NH2 -> \textcolor{Purple}{\ce{NO2}} + NH3} & -24.3 & \textbf{-24.0} & -10.0 \\
        \ce{\textcolor{Lavender}{\ce{HNO2}} + H -> \textcolor{Purple}{\ce{NO2}} + H2} & \textbf{-22.5} & -25.2 & -8.33 \\
        \noalign{\smallskip}
        \hline \hline
        \noalign{\smallskip}
        \ce{CHO + H2 -> \textcolor{CornflowerBlue}{\ce{CH2O}}} & \textbf{-6.35} & -9.17 & -2.14 \\
        \ce{CH2OH -> \textcolor{CornflowerBlue}{\ce{CH2O}} + H} & -11.8 & \textbf{-8.14} & -2.61 \\
        \noalign{\smallskip}
        \hline
        \noalign{\smallskip}
        \ce{HNO + CH3 -> \textcolor{Tan}{\ce{CH3NO}}}          & \textbf{-22.9} & \textbf{-21.3} & -8.62 \\
        \noalign{\smallskip}
        \hline \hline
        \noalign{\smallskip}
        \ce{\textcolor{Purple}{\ce{NO2}} + \textcolor{YellowOrange}{\ce{C2H5}} -> \textcolor{LimeGreen}{\ce{C2H5NO2}}} & -36.0 & -38.3 & \textbf{-18.3} \\
        \ce{\textcolor{Tan}{\ce{CH3NO}} + \textcolor{CornflowerBlue}{\ce{CH2O}} -> \textcolor{LimeGreen}{\ce{C2H5NO2}}} & \textbf{-35.2} & \textbf{-32.1} & -24.4 \\
		\noalign{\smallskip}
		\hline
	\end{tabular} \\
	\label{Tab_Gly_Pathways}
\end{table}

We find that all external energy sources lead to an enhancement of the glycine abundance in the upper atmosphere with the much higher flux of SEP and XUV changing the abundance by up to 18 orders of magnitude.
The lower CR flux only increases glycine by up to three orders of magnitude.
Nevertheless, the overall abundance of glycine is very low (below $10^{-30}$) such that the presence of glycine in the atmosphere of this planet will be challenging or impossiible to detect.
HCN, however, is much more abundant (up to $10^{-4}$) and similarly enhanced by SEP and XUV radiation.
Therefore, HCN provides an example of an observable signature that suggests the presence of prebiotic molecules.
CRs alone are not able to significantly enhance the abundance of HCN in the atmosphere.
This result presents an opportunity for the future study of the effect of lightning on the nightside of the planet:
with SEPs and XUV unable to reach that side of the planet, lightning might be the only way to significantly enhance the HCN abundance at the antistellar point.
We note, however, that this conclusion is only valid if horizontal transport between the day- and nightside can be neglected (Section~\ref{subsec_assym}).

Figure~\ref{HD189733b_abundances_lightning} shows the abundance of glycine and its precursors at the substellar (case 1) and antistellar (case 2) points with the same energy sources as in Fig.~\ref{HD189733b_abundances}, and the ratio between both abundances.
In addition, Fig.~\ref{Network_glycine} shows the reaction rates for the production pathways of glycine at the substellar point (SEP + XUV) at a pressure of $p = 2\times10^{-5}\si{\bar}$.
The numbers between the reactants and products in Fig.~\ref{Network_glycine} are the reaction rates ($\si{\per\centi\metre\cubed\per\second}$, $\log_{10}$) in the direction towards the sharp end of the lines.
The rates of the reverse reactions are not shown.
Table~\ref{Tab_Gly_Pathways} shows the rates of the most important production pathways of glycine for three different cases: the equilibrium substellar point (no SEP or XUV), the antistellar point (CR), and the substellar point (SEP and XUV).
The comparison of these different pathways shows that SEP and CR strongly enhance the production of \ce{C2H4} from \ce{C2H5^+} with the byproducts \ce{H3O^+} and \ce{NH4+} which belong to the most abundant ions in these scenarios (Fig.~\ref{HD189733b_abundances}).
At the substellar point (SEP + XUV), this pathway is the most efficient pathway of glycine production:
\begin{eqnarray}
    \ce{C2H5^+ + H2O &->& C2H4 + H3O^+} \\
    \ce{C2H4 + H &->& C2H5} \\
    \ce{C2H5 + NO2 &->& C2H5NO2}
\end{eqnarray}
Formaldehyde (\ce{CH2O}) and ethylene (\ce{C2H4}) belong to the most important and most abundant precursor molecules of glycine.
Similarly to glycine and HCN, they are strongly enhanced by the influx of SEP and XUV radiation in the upper atmosphere.
The bottom half of Fig.~\ref{HD189733b_glycine} shows the abundances of these organic molecules for different combinations of high-energy radiation.

\section{Conclusions}

In this work we studied the influence of different high-energy radiation sources on the atmospheric H/C/N/O chemistry of the hot Jupiter HD~189733b.
We combined recent simulations of the planetary atmosphere with the 3D Met Office Unified Model which self-consistently takes into account cloud formation \citep{Lines2018} with three different sources of high-energy radiation:
\begin{enumerate}
    \item multi-epochs observations of the stellar X-ray and UV spectrum \citep{Bourrier2020},
    \item low-energy cosmic rays \citep{Rimmer2013a}, and
    \item stellar energetic particles associated with an X-ray flare observed by \citet{Pillitteri2010} and based on a parametrization by \citet{Rab2017} and \citet{Herbst2019}.
\end{enumerate}
We then simulate the effect of these processes on the atmospheric chemistry of HD~189733b with the ion-neutral kinetics network \stand{}.

We identify ammonium (\ce{NH4^+}) and oxonium (\ce{H3O+}) as important and potentially observable signatures of an ionization of the atmosphere by cosmic rays and stellar particles.
Even though we note that glycine abundances remain low ($<10^{-30}$) in all our simulations, the influx of XUV radiation and SEPs enhances glycine by nearly 20 orders of magnitude.
In addition, we identify two important precursors for the production of glycine: formaldehyde (\ce{CH2O}) and ethylene (\ce{C2H4}), which are strongly enhanced by incoming XUV radiation and more so by the influx of SEPs.
Especially the abundance of ethylene gets enhanced to potentially observable values for \textit{JWST} in the upper atmosphere (Gasman et al., in prep.).
Ethylene has a strong absorption feature around $\SI{10}{\micro\metre}$ \citep{Mant2018} which could be detected by emission spectroscopy \citep{Hu2014}.
We therefore propose ethylene and potentially formaldehyde to be important signatures for incipient prebiotic synthesis under the influence of stellar energetic particles and XUV radiation.

Ethylene and other hydrocarbons have also been predicted by simulations of Jupiter's atmosphere by \citet{Moses2005} and with \standolder{} by \citet{Rimmer2016}.
In addition, these molecules were detected by observations in the stratosphere of Jupiter \citep[e.g.][]{Gladstone1996,Moses2005,Romani2008}.
Due to the very different temperature of Jupiter's atmosphere (in comparison to HD~189733b's atmosphere), however, we do not expect much glycine to form as most of the nitrogen and oxygen will be bound in \ce{NH3} and \ce{H2O} ice, respectively.

Even though hot Jupiters are not ideal places for the formation of complex prebiotic molecules such as glycine, they provide a unique laboratory to study the first steps of prebiotic synthesis under the influence of different high-energy radiation sources.
This is not least due to the better observability of giant gas planets in comparison to Earth-sized terrestrial planets.
Stellar energetic particles and XUV and FUV radiation strongly enhance the abundance of hydrocarbons and other organic molecules.
In particular, XUV radiation enhances the abundance of HCN to nearly $10^{-5}$ at the top of the atmosphere.
HCN is a crucial molecule in prebiotic chemistry and the formation of amino acids \citep{Airapetian2016,Airapetian2019}.
We note that the influx of cosmic rays on the nightside is not strong enough to produce additional HCN.
It will be therefore of special interest to study the production rate of HCN by lightning on the nightside where it could potentially be observable.

\section*{Acknowledgements}
We thank Konstantin Herbst for help with understanding and applying his peak size distribution to estimate the stellar proton flux and Stefan Lines for help with extracting the $pT$-profiles from the GCM output.
P.B. acknowledges a St Leonard's Interdisciplinary Doctoral Scholarship from the University of St Andrews.
Ch.H. is part of the CHAMELEON MC ITN EJD which received funding from the European Union’s Horizon 2020 research and innovation programme under the Marie Sklodowska-Curie grant agreement number 860470.
V.B. acknowledges support by the Swiss National Science Foundation (SNSF) in the frame of the National Centre for Competence in Research “PlanetS”, and has received funding from the European Research Council (ERC) under the European Union’s Horizon 2020 research and innovation programme (project Four Aces; grant agreement No 724427).
N.M. is partly supported by a Science and Technology Facilities Council Consolidated Grant (ST/R000395/1).
A.A.V. acknowledges funding from the ERC under the European Union's Horizon 2020 research and innovation programme (grant agreement No 817540, ASTROFLOW).
P.W. is supported by an STFC consolidated grant (ST/T000406/1).
R.F. acknowledges funding from UAEU startup grant number G00003269.

\section*{Data availability}
The data underlying this article were provided by the MOVES collaboration by permission. Data will be shared on request to the corresponding author with permission of the MOVES collaboration.
The version of the \stand{} network used in this article is available in the online supplementary material.



\bibliographystyle{mnras}
\bibliography{library_MOVESIV}




\bsp	
\label{lastpage}
\end{document}